\documentclass[preprint,superscriptaddress,showpacs,preprintnumbers,amsmath,amssymb,aps,showpacs,showkeys]{revtex4}

\usepackage[dvips]{epsfig,color}
\include{graphics}

\usepackage{graphicx}
\usepackage{dcolumn}

\begin{document}



\newcommand*{\FIU}{Florida International University, Miami, Florida 33199}
\affiliation{\FIU}
\newcommand*{\SCAROLINA}{University of South Carolina, Columbia, South 
Carolina 29208}
\affiliation{\SCAROLINA}
\newcommand*{\JLAB}{Thomas Jefferson National Accelerator Facility, Newport 
News, Virginia 23606}
\affiliation{\JLAB}
\newcommand*{\ANL}{Argonne National Laboratory, Argonne, Illinois 60439}
\affiliation{\ANL}
\newcommand*{\ASU}{Arizona State University, Tempe, Arizona 85287-1504}
\affiliation{\ASU}
\newcommand*{\UCLA}{University of California at Los Angeles, Los Angeles, 
California 90095-1547}
\affiliation{\UCLA}
\newcommand*{\CSU}{California State University Dominguez Hills, Carson,
California 90747}
\affiliation{\CSU}
\newcommand*{\CMU}{Carnegie Mellon University, Pittsburgh, Pennsylvania 15213}
\affiliation{\CMU}
\newcommand*{\CUA}{Catholic University of America, Washington, D.C. 20064}
\affiliation{\CUA}
\newcommand*{\SACLAY}{CEA-Saclay, Service de Physique Nucl\'eaire, 91191 
Gif-sur-Yvette, France}
\affiliation{\SACLAY}
\newcommand*{\CNU}{Christopher Newport University, Newport News, Virginia 
23606}
\affiliation{\CNU}
\newcommand*{\UCONN}{University of Connecticut, Storrs, Connecticut 06269}
\affiliation{\UCONN}
\newcommand*{\ECOSSEE}{Edinburgh University, Edinburgh EH9 3JZ, United Kingdom}
\affiliation{\ECOSSEE}
\newcommand*{\FU}{Fairfield University, Fairfield, Connecticut 06824}
\affiliation{\FU}
\newcommand*{\FSU}{Florida State University, Tallahassee, Florida 32306}
\affiliation{\FSU}
\newcommand*{\GEISSEN}{Physikalisches Institut der Universitaet Giessen, 
35392 Giessen, Germany}
\newcommand*{\GWU}{The George Washington University, Washington, DC 20052}
\affiliation{\GWU}
\newcommand*{\ECOSSEG}{University of Glasgow, Glasgow G12 8QQ, United Kingdom}
\affiliation{\ECOSSEG}
\newcommand*{\ISU}{Idaho State University, Pocatello, Idaho 83209}
\affiliation{\ISU}
\newcommand*{\INFNFR}{INFN, Laboratori Nazionali di Frascati, 00044 Frascati, 
Italy}
\affiliation{\INFNFR}
\newcommand*{\INFNGE}{INFN, Sezione di Genova, 16146 Genova, Italy}
\affiliation{\INFNGE}
\newcommand*{\ORSAY}{Institut de Physique Nucleaire ORSAY, Orsay, France}
\affiliation{\ORSAY}
\newcommand*{\ITEP}{Institute of Theoretical and Experimental Physics, Moscow,
117259, Russia}
\affiliation{\ITEP}
\newcommand*{\JMU}{James Madison University, Harrisonburg, Virginia 22807}
\affiliation{\JMU}
\newcommand*{\KYUNGPOOK}{Kyungpook National University, Daegu 702-701, Republic of Korea}
\affiliation{\KYUNGPOOK}
\newcommand*{\MIT}{Massachusetts Institute of Technology, Cambridge, 
Massachusetts 02139-4307}
\affiliation{\MIT}
\newcommand*{\UMASS}{University of Massachusetts, Amherst, Massachusetts  
01003}
\affiliation{\UMASS}
\newcommand*{\MOSCOW}{Moscow State University, General Nuclear Physics 
Institute, 119899 Moscow, Russia}
\affiliation{\MOSCOW}
\newcommand*{\UNH}{University of New Hampshire, Durham, New Hampshire 
03824-3568}
\affiliation{\UNH}
\newcommand*{\NSU}{Norfolk State University, Norfolk, Virginia 23504}
\affiliation{\NSU}
\newcommand*{\OHIOU}{Ohio University, Athens, Ohio 45701}
\affiliation{\OHIOU}
\newcommand*{\ODU}{Old Dominion University, Norfolk, Virginia 23529}
\affiliation{\ODU}
\newcommand*{\PITT}{University of Pittsburgh, Pittsburgh, Pennsylvania 15260}
\affiliation{\PITT}
\newcommand*{\RPI}{Rensselaer Polytechnic Institute, Troy, New York 12180-3590}
\affiliation{\RPI}
\newcommand*{\RICE}{Rice University, Houston, Texas 77005-1892}
\affiliation{\RICE}
\newcommand*{\URICH}{University of Richmond, Richmond, Virginia 23173}
\affiliation{\URICH}
\newcommand*{\UNIONC}{Union College, Schenectady, NY 12308}
\affiliation{\UNIONC}
\newcommand*{\VT}{Virginia Polytechnic Institute and State University, 
Blacksburg, Virginia 24061-0435}
\affiliation{\VT}
\newcommand*{\VIRGINIA}{University of Virginia, Charlottesville, Virginia 
22901}
\affiliation{\VIRGINIA}
\newcommand*{\WM}{College of William and Mary, Williamsburg, Virginia 
23187-8795}
\affiliation{\WM}
\newcommand*{\YEREVAN}{Yerevan Physics Institute, 375036 Yerevan, Armenia}
\affiliation{\YEREVAN}


\author {R.~Nasseripour} \affiliation{\FIU}\affiliation{\SCAROLINA}
\author {B.A.~Raue} \affiliation{\FIU}
\author {P.~Ambrozewicz} \affiliation{\FIU}
\author {D.S.~Carman} \affiliation{\JLAB}
\author {M.J.~Amaryan} \affiliation{\ODU}
\author {E.~Anciant} \affiliation{\SACLAY}
\author {M.~Anghinolfi} \affiliation{\INFNGE}
\author {B.~Asavapibhop} \affiliation{\UMASS}
\author {G.~Asryan} \affiliation{\YEREVAN}
\author {G.~Audit} \affiliation{\SACLAY}
% ...
\collaboration{The CLAS Collaboration}
     \noaffiliation

\title{Polarized Structure Function $\sigma_{LT'}$ for 
$p({\vec e},e'K^+)\Lambda$ in the Nucleon Resonance Region}
\date{\today}

\begin{abstract}
The first measurements of the polarized structure function $\sigma_{LT'}$ 
for the reaction $p(\vec e,e'K^+)\Lambda$ in the nucleon resonance region
are reported.  Measurements are included from threshold up to $W$=2.05~GeV 
for central values of $Q^2$ of 0.65 and 1.00~GeV$^2$, and nearly the 
entire kaon center-of-mass angular range.  $\sigma_{LT'}$ is the imaginary 
part of the longitudinal-transverse response and is expected to be sensitive 
to interferences between competing intermediate $s$-channel resonances, as 
well as resonant and non-resonant processes.  The results for $\sigma_{LT'}$ 
are comparable in magnitude to previously reported results from CLAS for 
$\sigma_{LT}$, the real part of the same response.  An intriguing sign 
change in $\sigma_{LT'}$ is observed in the high $Q^2$ data at 
$W\approx 1.9$~GeV.  Comparisons to several existing model predictions are 
shown.
\end{abstract}


\maketitle


\section{Introduction}
\label{sec:intro}

The study of the electromagnetic production of strange quarks in the
resonance region plays an important role in understanding the strong
interaction.  The $p(e,e'K^+)\Lambda$ reaction involves the production of
the strange particles $\Lambda$($uds$) and $K^+$($u\bar{s}$) in the final
state via strange quark-pair ($s\bar{s}$) creation.  The fundamental
theory for the description of the dynamics of quarks and gluons is known
as quantum chromodynamics (QCD).  However, while numerical approaches to 
QCD in the medium-energy regime do exist, neither perturbative QCD nor 
lattice QCD can presently predict hadron properties seen in this type of 
reaction.  In the non-perturbative regime of nucleon resonance physics, 
the consequence is that the interpretation of dynamical hadronic processes 
still hinges to a significant degree on models containing some 
phenomenological ingredients.  Various quark models (see for example 
Refs.~\cite{isgur,capstick1,capstick2,capstick3}) predict a large number of 
non-strange baryons that can decay into a strange baryon and a strange meson, 
as well as N$\pi$/N$\pi\pi$ final states.  While many
of these excited states have been observed in pion production data, a large
number are ``missing''.  The higher threshold for $K^+\Lambda$ final states
kinematically favors production of the missing resonances with masses near
2~GeV.  Studies of different final states, such as the associated
production of strangeness, can provide complementary information on the
contributing amplitudes.

\begin{figure}[htpb]
\vspace{5.2cm}
\includegraphics{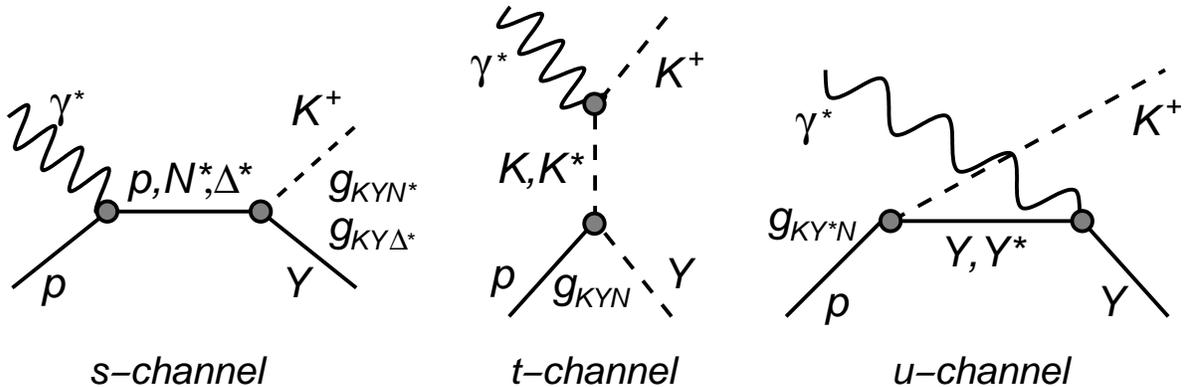}
\caption{Feynman diagrams representing $s$-channel nucleon 
($p,N^*,\Delta^*$) exchange (left), $t$-channel kaon ($K,K^*$) 
exchange (middle), and $u$-channel hyperon ($Y,Y^*$) exchange (right) 
that contribute to the reaction models.  The vertex labels $g_{MBB}$ 
represent the strong coupling constants.}
\label{fig:fig-chan}
\end{figure}

In the absence of direct QCD predictions, effective models must be
employed.  Utilizing these models by means of fitting them to the available
experimental data -- cross sections and polarization observables -- or
comparing the data to the model predictions, can provide information on the
reaction dynamics.  In addition, these comparisons can provide important
qualitative and quantitative information on the contributing resonant and
non-resonant terms in the $s$, $t$, and $u$ reaction channels (see
Fig.~\ref{fig:fig-chan}).  The development of these theoretical models has
been highly based on the availability of the experimental data.  Precise
measurements of cross sections and polarization observables are crucial for
the refinement of these models and for the search for missing resonances.

In this paper, we report the first-ever measurements of the 
longitudinal-transverse polarized structure function, $\sigma_{LT'}$, 
for the $p(\vec{e},e'K^+)\Lambda$ reaction in the resonance region, using 
the CEBAF Large Acceptance Spectrometer (CLAS) in Hall B of Jefferson Lab.  
This observable provides complementary information to the $\sigma_{LT}$ 
structure function reported in Ref.~\cite{5st}, as will be discussed.  Thus,
these new data provide another constraint on model parameters, and
therefore, provide additional important information in understanding the 
process of electromagnetic production of strangeness.

There is a growing body of high-quality data on the electromagnetic 
production of strange hadrons.  Recently published data using electron
beams exist on the separation of the longitudinal and transverse structure 
functions, $\sigma_L$ and $\sigma_T$, from Hall C of Jefferson Lab for both 
$K^+\Lambda$ and $K^+\Sigma^0$ final states at $W$=1.84~GeV, for $Q^2$ 
up to 2.0~GeV$^2$, at a kaon center-of-mass scattering angle of 
$\theta_K^*=0^\circ$~\cite{gabriel,mohring}.  The CLAS Collaboration has 
recently produced results in which unpolarized cross sections and 
interference structure functions ($\sigma_{TT}$ and $\sigma_{LT}$) have 
been measured for $K^+\Lambda$ and $K^+\Sigma^0$ final states over a wide 
kinematic range with $Q^2$ up to 2.6~GeV$^2$, $W$ up to 2.4~GeV, and nearly 
complete angular coverage in the center-of-mass frame~\cite{5st}.  These 
results include the first-ever separation of $\sigma_L$ and $\sigma_T$ at 
angles other than $\theta_K^*=0^\circ$.  The same set of data has been 
analyzed to extract the polarization transfer from the virtual photon to 
the produced $\Lambda$ hyperon~\cite{carman} and to extract the ratio of 
$\sigma_L/\sigma_T$ at $\theta_K^*=0^\circ$ for the $K^+\Lambda$ final 
state~\cite{raue05}.  Older electroproduction data from various labs also 
exist~\cite{brown,bebek1,azemoon,bebek2,brauel}, but with much larger 
uncertainties and much smaller kinematic coverage.

Complementary data from photoproduction are also available.  The SAPHIR 
collaboration has published total and differential cross section data for 
photoproduction of $K^+\Lambda$ and $K^+\Sigma^0$ final states with photon 
energies up to 2~GeV \cite{saphir1,saphir2}.  CLAS has provided extensive 
differential cross sections~\cite{mcnabb,bradford1}, along with recoil
\cite{mcnabb} and transferred polarization~\cite{bradford2} data for
the same final states in similar kinematics.  Finally, the LEPS collaboration 
has measured differential cross sections and polarized beam asymmetries with 
a linearly polarized photon beam for energies up to 2.4~GeV at forward angles
\cite{zegers,sumihama}.

The SAPHIR cross section data show an interesting resonance-like structure 
in the $K^+\Lambda$ final state around $W$=1.9~GeV.  A similar structure 
has been seen in the unpolarized electroproduction cross section data
\cite{5st}, as well as in the photoproduction measurements of CLAS
\cite{mcnabb,bradford1}.   Within the isobar model of Mart and Bennhold
\cite{d13mart,mart}, that structure was interpreted as a $D_{13}(1895)$ 
resonance, which had been predicted by several quark models (e.g. 
Ref.~\cite{capstick3}), but not well established.  However, the isobar 
model of Saghai~\cite{saghai} found that the cross section data could be 
satisfactorily described without the need for including any new $s$-channel 
resonances by including higher-spin $u$-channel exchange terms.  The need 
to include the missing $D_{13}(1895)$ state, however, was supported by 
the new Regge plus resonance model from Ghent~\cite{ghent} that 
compared their model to a broad set of cross section and polarization 
observables from the available photo- and electroproduction data.  

The organization of this paper includes an overview of the relevant
formalism in Section~\ref{formal}, a description of the theoretical
models used to compare against the data in Section~\ref{sec:theory},
details on the experiment and data analysis in Section~\ref{sec:exper}, 
and a presentation and discussion of the results in Section~\ref{sec:result}.  
The conclusions are given in Section~\ref{sec:conc}.

	
\section{Formalism}
\label{formal}

\begin{figure}[tbp]
\begin{center}
\mbox{\epsfxsize=8.0cm\leavevmode \epsffile{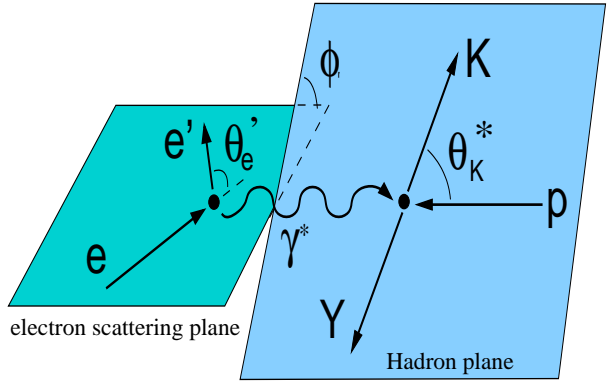}}	
\end{center}
\caption{(Color online) Kinematic diagram for kaon-hyperon ($KY$) 
electroproduction.}
\label{fig:kin}
\end{figure}

A schematic diagram of $K^+\Lambda$ electroproduction  off a fixed 
hydrogen target is shown in Fig.~\ref{fig:kin}.  The angle between 
the incident and scattered electron is $\theta_e$, while the angle 
between the electron scattering plane and hadron production plane 
is defined as $\phi$.  In the one-photon exchange approximation, the 
interaction between the incident electron beam and the target proton 
is mediated by a virtual photon, $\gamma^*$.  The virtual photon four 
momentum is obtained from the difference between the four momenta of 
the incident, $e = (E, \mathbf p_e \rm)$, and scattered electrons, 
$e' = (E', \mathbf p'_e \rm)$, as:
\begin{equation}
q = e - e' = (\nu, \mathbf q).
\end{equation}
The four momentum transfer squared, $Q^2$, is an invariant quantity
defined as:
\begin{equation}
Q^2 = -q^2 = -(\nu^2 - \mathbf q^2) = 4EE' \sin^2(\theta_{e'}/2),
\end{equation}
where $\nu = E - E'$ is the energy transfer and $\theta_{e'}$ is the
electron scattering angle in the lab frame.  The invariant mass $W$ of 
the intermediate hadronic state is defined as:
\begin{equation}
W^2 = s = M^2 + 2M\nu - Q^2,
\end{equation}
where $M$ is the mass of the proton target.

Following the notation of Refs.~\cite{akerlof,zeit}, the differential 
cross section for $KY$ electroproduction in the center-of-mass frame is 
given by:
\begin{equation}
\frac {d\sigma}{dE' d\Omega_e' d\Omega_K^*} = \Gamma 
\frac{d\sigma}{d\Omega_K^*},
\end{equation}
where $\Gamma$ is the virtual photon flux given by:
\begin{equation}
\label{Gamma}
 \Gamma = \frac{\alpha}{4\pi^2}\frac{E'}{E M}
\frac{W^2-M^2}{Q^2} \left( \frac{1}{1-\epsilon} \right).
\end{equation}
Here $\frac {d\sigma}{d\Omega_K^*}$ is the virtual photon cross
section and $\epsilon$ is the virtual photon transverse polarization 
component defined as:
\begin{equation}
\epsilon = \left( 1+ 2 \left( 1 + \frac{\nu^2}{Q^2} \right) 
\tan^2\frac{\theta_{e'}}{2}\right)^{-1}.
\end{equation}

The cross section for the electromagnetic interaction of a relativistic
electron beam with a hadron target is obtained by calculating the
transition probability of the process~\cite{boffi}. The cross section can 
be written in the form of a contraction between leptonic and hadronic 
tensors that contain the electron and hadron variables separately. In 
general, the lepton tensor can be written in terms of a density matrix of 
virtual photon polarization that contains a symmetric helicity-independent 
part and an anti-symmetric helicity-dependent part. The anti-symmetric part 
contributes to the cross section only when the hadron tensor also contains an
anti-symmetric part.  This is the case when scattering a polarized electron
off of an unpolarized target with the detection of the final state hadron in
coincidence with the scattered electron. The anti-symmetric part vanishes 
for the case of unpolarized electrons. 

For a polarized electron beam with helicity $h$ and no target or
recoil polarizations, the virtual photon cross section can be
written as:
\begin{equation}
\frac{d\sigma}{d\Omega_K^*} = \sigma_T + \epsilon \sigma_L
+ \epsilon \sigma_{TT} \cos 2\phi + \sqrt{\epsilon(1 + \epsilon)}
\sigma_{LT} \cos\phi + h \sqrt{\epsilon(1 - \epsilon)} \sigma_{LT'} 
\sin\phi, 
\label{eq:full_csec}
\end{equation} 
where $\sigma_i$ are the structure functions that measure the response 
of the hadronic system and $i=T$, $L$, $LT$, $TT$, and $LT'$ represent 
the transverse, longitudinal, and interference structure functions.
The structure functions are, in general, functions of $Q^2$, $W$, 
and $\theta_K^*$ only.  Note that the convention employed here for the 
differential cross section is not used by all authors~\cite{convention}.

For the case of an unpolarized electron beam, Eq.(\ref{eq:full_csec}) 
reduces to the unpolarized cross section, $\sigma_0$:
\begin{equation}
\label{sigma0}
\frac{d\sigma}{d\Omega_K^*} \equiv \sigma_0 = \sigma_T + 
\epsilon \sigma_L + \epsilon \sigma_{TT} \cos 2\phi + 
\sqrt{\epsilon(1+\epsilon)} \sigma_{LT} \cos \phi.
\end{equation}
The electron polarization therefore produces a fifth structure 
function that is related to the beam helicity asymmetry via:
\begin{equation}
\label{eq:sigltp}
A_{LT'} = \frac{\frac{d\sigma}{d\Omega_K^*}^+ - \frac{d\sigma}{d\Omega_K^*}^-}
{\frac{d\sigma}{d\Omega_K^*}^+ + \frac{d\sigma}{d\Omega_K^*}^-} = 
\frac{\sqrt{\epsilon(1-\epsilon)}\sigma_{LT'} \sin\phi}{\sigma_0}.
\end{equation}
The $\pm$ superscripts on $\frac{d\sigma}{d\Omega_K^*}$ correspond to the
electron helicity states of $h=\pm 1$.  Clearly, $\sigma_{LT'}$ can only be
observed when the outgoing hadron is detected out of the electron
scattering plane ($\phi \neq 0$) and can be separated by flipping the
electron helicity.

The structure functions are defined in terms of the independent elements 
of the hadron tensor in the center-of-mass frame, $W'_{\lambda\lambda'}$
\cite{boffi}:
\begin{eqnarray}
\label{eq:ampsltp}
\sigma_L & \propto & W^{'}_{00}, \nonumber \\ 
\sigma_T  & \propto  & (W^{'}_{11} + W^{'}_{-1-1}), \nonumber \\
\sigma_{TT}  & \propto  & W^{'}_{1-1}, \\
\sigma_{LT}  & \propto  & Re (W^{'}_{01} - W^{'}_{0-1}), \nonumber \\
\sigma_{LT'}  & \propto  & Im (W^{'}_{01} - W^{'}_{0-1}), \nonumber 
\end{eqnarray}
where the indices $\lambda,\lambda' = 0$ for the longitudinal component 
and $\lambda,\lambda' =\pm 1$ for the two transverse components.  In
contrast to the case of real photons, where there is only the purely
transverse response, virtual photons allow longitudinal, 
transverse-transverse, and longitudinal-transverse interference terms to 
occur. 

The polarized structure function $\sigma_{LT'}$ is intrinsically different
from the four structure functions of the unpolarized cross section.  As 
seen by Eqs.(\ref{eq:ampsltp}), this term is generated by the imaginary 
part of terms involving the interference between longitudinal and transverse
components of the hadronic and leptonic currents.  This is in contrast to 
$\sigma_{LT}$, which is generated by the real part of the same interference.  
$\sigma_{LT'}$ is non-vanishing only if the hadronic tensor is anti-symmetric,
which will occur in the presence of final state interaction (FSI) (or
rescattering) effects, interferences between multiple resonances, 
interferences between resonant and non-resonant processes, or even between 
non-resonant process alone.  On the other hand, $\sigma_{LT'}$ could be 
non-zero even when $\sigma_{LT}$ (which is not expected to be sensitive to 
FSI effects~\cite{boffi}) is zero. It provides a means of measuring the 
contributions of small resonance channels that are often too weak to be 
observed directly in the unpolarized cross sections.  Furthermore, when 
the reaction proceeds through a channel in which a single amplitude 
dominates, the longitudinal-transverse response will be real and 
$\sigma_{LT'}$ vanishes.  Both $\sigma_{LT}$ and $\sigma_{LT'}$ are necessary 
to fully unravel the longitudinal-transverse response of the $K^+\Lambda$ 
electroproduction reaction.


\section{Theoretical Models}
\label{sec:theory}

With the recently available data from the photo- and electroproduction of
$KY$ final states from CLAS and elsewhere, there have been renewed efforts
on the development of theoretical models.  The majority of these are
single-channel models that represent tree-level calculations, where the 
amplitude is constructed from the lowest-order Feynman diagrams (see 
Ref.~\cite{ghent} and references therein).  More recent work has moved 
beyond the single-channel approach with the development of coupled-channels 
models~\cite{juliadiaz,chiang,shklyar,leecole} or by fitting simultaneously 
to multiple, independent reaction channels~\cite{sarantsev,anisov}.  However, 
as a combined coupled-channels analysis of the photo- and electroproduction 
reactions is not yet available, a tree-level approach currently represents 
the best possibility of studying both reactions within the same framework.  
While most of the recent theoretical analyses have focused solely on the 
available photoproduction data, it has been shown that electroproduction 
observables can yield important complementary insights to improve and 
constrain theory~\cite{ghent}.

At the medium energies used in this experiment, perturbative QCD is not 
capable of providing any predictions for the differential cross 
sections or structure functions for kaon electroproduction.  In this work, 
the results are compared against three different model approaches.  The first 
is a traditional hadrodynamic (resonance) model,  the second is based on a 
Reggeon-exchange model, and the third is a hybrid Regge plus resonance 
approach.  

In the hadrodynamic model approach, the strong interaction is modeled by 
an effective Lagrangian, which is constructed from tree-level Born and 
extended Born terms for intermediate states exchanged in the $s$, $t$, and 
$u$ reaction channels (see Fig.~\ref{fig:fig-chan}).  Each resonance has 
its own strong coupling constants and strong decay widths.  A complete 
description of the physics processes requires taking into account all 
possible channels that could couple to the initial and final state measured, 
but the advantages of the tree-level approach include the ability to limit 
complexity and to identify the dominant trends.  In the one-channel, 
tree-level approach, several dozen parameters must be fixed by fitting to 
the data, since they are poorly known and not constrained from other sources.  

The hadrodynamic model employed in this work was developed by Mart and 
Bennhold ~\cite{mart,lee} (referred to here as MB).  In this model, the 
coupling strengths have been determined mainly by fits to existing 
$\gamma p \to K^+ Y$ data (with some older electroproduction data included), 
leaving the coupling constants as free parameters (constrained loosely by 
SU(3) symmetry requirements).  It employs phenomenological form factors to 
account for the extension of the point-like interactions at the hadronic 
vertices.  This model has been compared against the existing photoproduction 
data from SAPHIR~\cite{saphir1,saphir2} and CLAS~\cite{mcnabb,bradford1}, and 
provides a fair description of those results.  The model parameters are not 
based on fits to any CLAS data.  The specific resonances included in this 
model are the $S_{11}$(1650), $P_{11}$(1710), $P_{13}$(1720), and 
$D_{13}$(1895) $N^*$ states in the $s$-channel, and the $K^*$(892) and 
$K^*_1$(1270) in the $t$-channel.

The data are also compared to the Reggeon-exchange model from Guidal, Laget, 
and Vanderhaeghen~\cite{guidal} (referred to here as GLV).  This calculation 
includes no baryon resonance terms at all.  Instead, it is based only on 
gauge-invariant $t$-channel $K$ and $K^*$ Regge-trajectory exchange.  It 
therefore provides a complementary basis for studying the underlying 
dynamics of strangeness production.  It is important to note that the Regge 
approach has far fewer parameters compared to the hadrodynamic models.  
These include the $K$ and $K^*$ form factors (assumed to be of a monopole 
form) and the coupling constants $g_{KYN}$ and $g_{K^*YN}$ (taken from 
photoproduction studies).

The GLV model was fit to higher-energy photoproduction data where there is 
little doubt of the dominance of these kaon exchanges, and extrapolated 
down to JLab energies.  An important feature of this model is the way 
gauge invariance is achieved for the $K$ and $K^*$ $t$-channel exchanges 
by Reggeizing the $s$-channel nucleon pole contribution in the same manner 
as the $t$-channel diagrams~\cite{guidal}.  Due to gauge invariance, the 
$t$-channel exchanges and $s$-channel nucleon pole terms are inseparable 
and are treated on the same footing.  They are Reggeized in the same way 
and multiplied by the same electromagnetic form factor.  No counter terms 
need to be introduced to restore gauge invariance as is done in the 
hadrodynamic approach.

The final model included in this work was developed by the University of
Ghent group~\cite{ghent}, and is based on a tree-level effective field model 
for $\Lambda$ and $\Sigma^0$ photoproduction from the proton.  It differs 
from traditional isobar approaches in its description of the non-resonant 
diagrams, which involve the exchange of $K$ and $K^*$ Regge trajectories.  
A selection of $s$-channel resonances are then added to this background.  
This ``Regge plus resonance'' (referred to here as RPR) approach has the 
advantage that the background diagrams contain only a few parameters that 
are constrained by high-energy data where the $t$-channel processes dominate.  
Furthermore, the use of Regge propagators eliminates the need to introduce 
strong form factors in the background terms, thus avoiding the 
gauge-invariance issues associated with the traditional effective Lagrangian 
models.  In addition to the kaonic trajectories to model the $t$-channel 
background, the RPR model includes the $s$-channel resonances $S_{11}$(1650), 
$P_{11}$(1710), $P_{13}$(1720), and $P_{13}$(1900).  Apart from these, the 
model includes either a $D_{13}$(1900) or $P_{11}$(1900) state in the 
$K^+\Lambda$ channel.  In detailed comparisons with the separated structure 
functions~\cite{5st} and beam-recoil transferred polarization data from 
CLAS~\cite{carman}, only the $D_{13}$(1900) assumption could be reconciled 
with the data, whereas the $P_{11}$(1900) option could clearly be rejected
\cite{ghent}.  Note that the CLAS electroproduction data~\cite{5st} strongly 
suggest a reaction mechanism for $K^+\Lambda$ dominated by $t$-channel 
exchange, however there are obvious discrepancies with the Regge predictions, 
indicative of $s$-channel contributions.


\section{Experiment and Data Analysis}
\label{sec:exper}

\subsection{Experimental Apparatus}

The data included in this work were taken in 1999, using the high duty 
factor electron beam at Jefferson Lab and the CEBAF Large Acceptance 
Spectrometer (CLAS)~\cite{clas} in Hall B.  A longitudinally polarized 
2.567~GeV electron beam with a current of 5~nA was incident upon a 5-cm-long 
liquid-hydrogen target with a density of 0.073~g/cm$^3$, resulting in a 
luminosity of $\sim 10^{34}$~cm$^{-2}$s$^{-1}$.  The electron beam 
polarization was measured regularly throughout the experiment with a 
coincidence M{\o}ller polarimeter~\cite{clas}.  The average beam 
polarization was measured to be 67.0$\pm$1.5\%.

CLAS is a large acceptance spectrometer used to detect multi-particle 
final states.  Six superconducting coils generate a toroidal magnetic 
field around the target with azimuthal symmetry about the beam axis.
The coils divide CLAS into six sectors, each functioning as an 
independent magnetic spectrometer.  Each sector is instrumented with 
drift chambers (DC) to determine charged-particle trajectories~\cite{dc}, 
scintillator counters (SC) for time-of-flight measurements~\cite{sc}, and, 
in the forward region, gas-filled threshold \v{C}erenkov counters (CC) for 
electron/pion separation up to 2.5~GeV~\cite{cc} and electromagnetic 
calorimeters (EC) to identify and measure the energy of electrons and 
high-energy neutral particles, as well as to provide electron/pion 
separation above 2.5~GeV~\cite{ec}.  The trigger for the data acquisition 
readout of CLAS was a coincidence between the CC and EC in a given sector, 
which selected the electron candidates.  For the data sets used in the 
present work, the total number of triggers collected was 530~M and 370~M 
for the two torus current settings of 1500~A and 2250~A, respectively.  
These two data sets were combined together for the present analysis.


\subsection{Data Binning}

The data were binned in a four-dimensional space of the independent
kinematic variables, $Q^2$, $W$, $\cos \theta_K^*$, and $\phi$.
Table~\ref{tab:bins} gives the binning in the variables $Q^2$, $W$, 
and $\cos \theta_K^*$, while $\phi$ was binned in eight, equal-sized 
bins running from -180$^\circ$ to 180$^\circ$.  A small fraction 
($< 5\%$) of the $\phi$ bins have been excluded from this analysis due 
to their low acceptance in CLAS.  A point was rejected if its acceptance 
was less than 2.0\% (absolute) or less than 10\% of the average acceptance 
over all bins at the same $Q^2$, $W$, and $\cos\theta_K^*$.  These tend to 
be the bins adjacent to $\phi=0^\circ$ where the asymmetry is small because 
of the $\sin\phi$ dependence seen in Eq.(\ref{eq:sigltp}), and therefore 
their absence has little effect on the extraction of $\sigma_{LT'}$.

\begin{table}[htbp]
\begin{center}
\begin{tabular} {|c|c|c|c|c|c|} \hline
\multicolumn{2}{|c|}{$Q^2$ (GeV$^2$)} & \multicolumn{2}{|c|}{$W$ (GeV)}
& \multicolumn{2}{|c|}{$\cos\theta_K^*$}\\ \hline
Range & Bin Center & Range & Bin Center & Range & Bin Center \\ \hline
0.50 to 0.80 & 0.65 & 1.60 to 1.70 & 1.650 & -0.80 to -0.40 & -0.60~ \\
0.80 to 1.30 & 1.00 & 1.70 to 1.75 & 1.725 & -0.40 to -0.10 & -0.25~ \\
                   && 1.75 to 1.80 & 1.775 & -0.10 to  0.20 &  0.05 \\
                   && 1.80 to 1.85 & 1.825 &  0.20 to  0.50 &  0.35 \\
                   && 1.85 to 1.90 & 1.875 &  0.50 to  0.80 &  0.65 \\
                   && 1.90 to 1.95 & 1.925 &  0.80 to  1.00 &  0.90 \\
                   && 1.95 to 2.00 & 1.975 &                &       \\
                   && 2.00 to 2.10 & 2.050 &                &       \\ \hline
\hline
\end{tabular}
\caption{Ranges and centers of the kinematic bins used in this analysis.
Note that the $Q^2$ bin from 0.8 to 1.3~GeV$^2$ was bin centered to the
value of 1.00~GeV$^2$ in this work (see Section~\ref{sec:syserr}).}
\label{tab:bins}
\end{center}
\end{table}

\subsection{Particle Identification}

The $p(e,e'K^+)\Lambda$ reaction was isolated by detecting the scattered 
electron, $e'$, and kaon, $K^+$, with CLAS, and reconstructing the
hyperon via the missing mass technique.  Electrons were identified by 
producing an electromagnetic shower in the EC accompanied by a signal 
in the CC.  The electron energy deposited in the EC for all electron 
candidates was required to be consistent with the momentum measured by 
the track reconstruction in the DC.  Electron and pion separation was also 
made by distinguishing between their different interaction modes in the EC.  
The start time of the interaction was then obtained by calculating the 
difference between the time measured by the SC and the flight time measured 
by the DC.  This measured start time was combined with the hadron momentum 
and the path length measured by the DC to determine the hadron mass.  

Corrections to the electron and kaon momenta were devised to correct for
reconstruction inaccuracies.  These arise from the relative misalignments of 
the drift chambers in the CLAS magnetic field, as well as for uncertainties 
in the magnetic field map employed during charged track reconstructions.  
These corrections were typically less than 1\%.

\begin{figure}[htbp]
\vspace{9.2cm}
\includegraphics{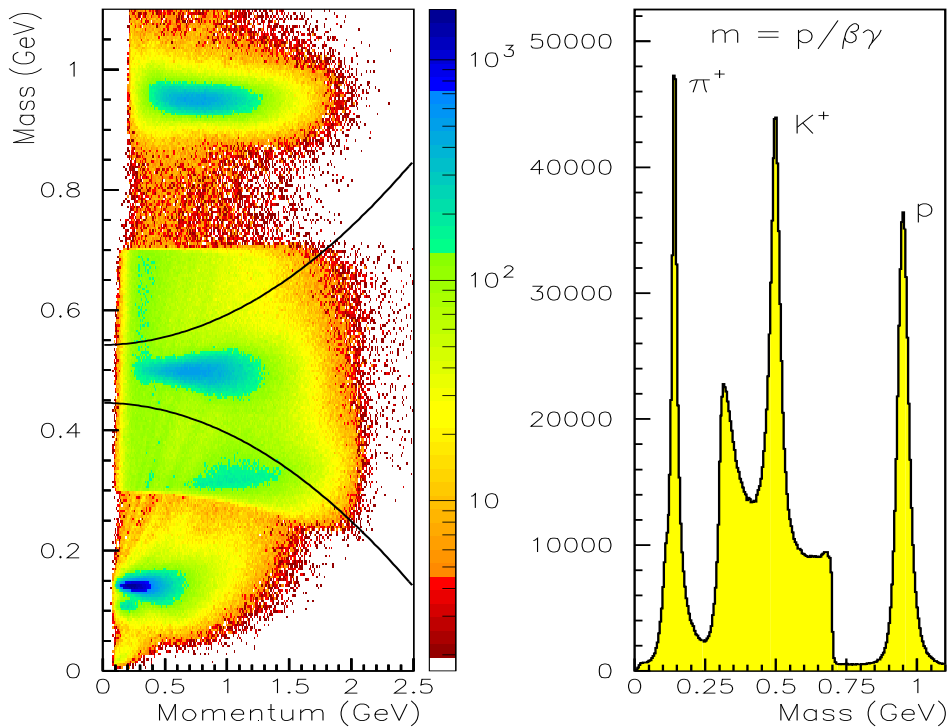}
\caption{(Color online) Reconstructed mass for positively 
charged particles.  The left figure shows the mass plotted against 
the measured momentum. The lines show the mass cuts used to identify 
kaon candidates.  A logarithmic yield density scale is employed.  The 
right figure shows the reconstructed mass.  These spectra were made 
from the kaon-filtered data files. The kaon peak is enhanced relative
to the pion and proton peaks in the range from 0.3 to 0.7~GeV due to
the filtering condition.}
\label{kmass}
\end{figure}

Due to the small fraction of events containing kaons in the CLAS 
data, a pre-selection of kaon events based on preliminary particle 
identification was made. Here the kaon candidates were selected by 
choosing positively charged particles with a reconstructed mass between 
0.3 and 0.7~GeV.  Because the relative momentum resolution of CLAS becomes 
poorer with increasing momentum, a momentum-dependent mass cut was used.  
Fig.~\ref{kmass}a shows the reconstructed hadron mass as a function of 
momentum along with the cut used.  Fig.~\ref{kmass}b shows the projected 
hadron mass distribution for all hadrons that passed the pre-selection 
criterion.

\begin{figure}[htbp]
\vspace{9.2cm}
\includegraphics{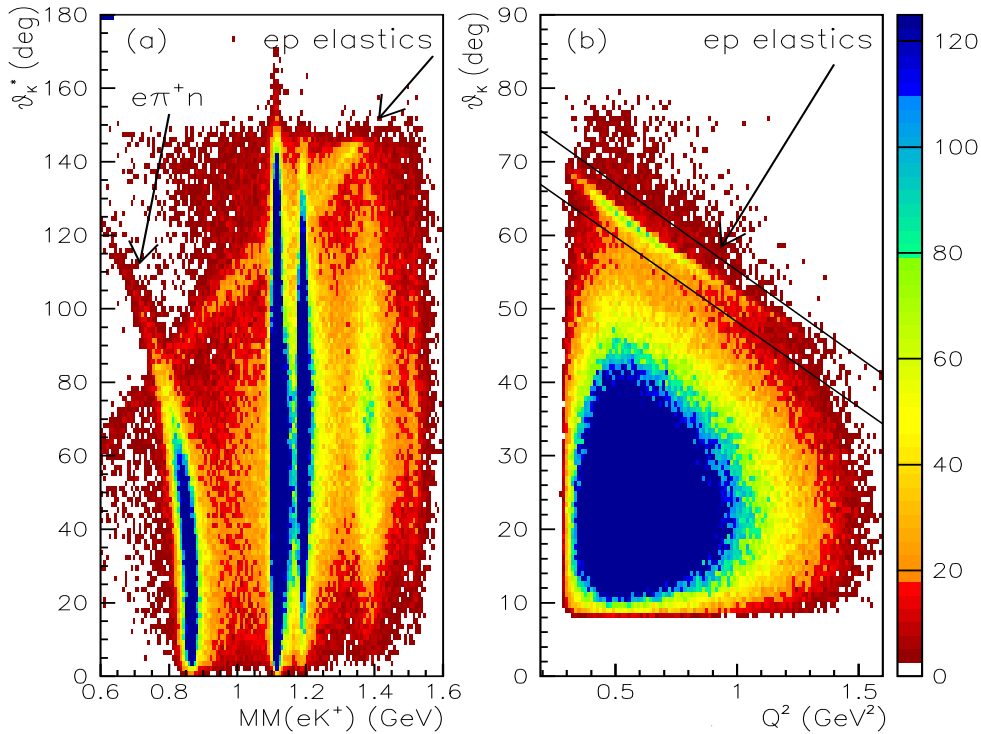}
\caption{(Color online) (a) $\theta_K^*$ vs. $p(e,e'K^+)$ missing 
mass showing $ep$ elastic events and $e'\pi^+ n$ events.  The vertical 
bands correspond to ground state $\Lambda(1116)$ and $\Sigma^0(1193)$
hyperons, and the $\Sigma^0(1385)/\Lambda(1405)$ hyperons. (b) $\theta_K$ 
(lab angle) vs. $Q^2$ for $p(e,e'K^+)$ events showing the $ep$ elastic 
events and the cut used to remove them.}
\label{fig-elastics}
\end{figure}

\begin{figure}[htpb]
\vspace{6.5cm}
\includegraphics{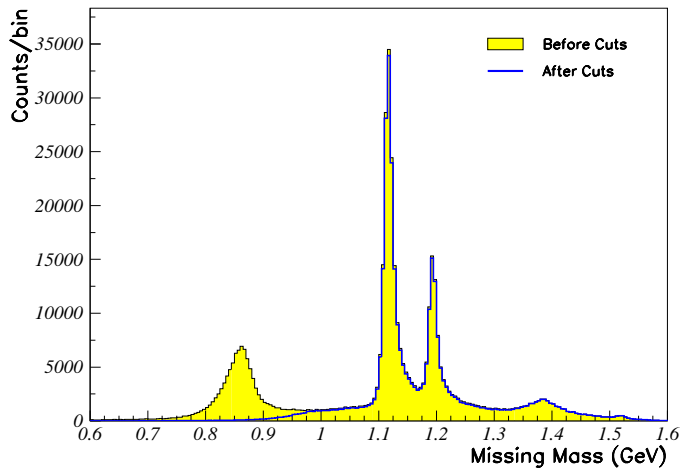}
\caption{(Color Online) Missing mass for $p(e,e'K^+)$ summed over 
the entire range of $Q^2$, $W$, $\cos\theta_K^*$, and $\phi$ of the 
data before and after removing the elastic $ep$ events and the 
$e'\pi^+ n$ events.}
\label{mmplot}
\end{figure}

Hyperons are identified by using the four-momenta of the electron 
beam, scattered electron, and the $K^+$ candidate.  The missing mass 
distribution contains a background that includes a continuum beneath 
the hyperons from multi-particle final states with misidentified 
pions and protons, as well as events from $ep$ elastic scattering 
(protons misidentified as kaons) and events from $\pi^+ n$ final 
states (pions misidentified as kaons).  The elastic events are 
kinematically correlated and show up clearly in plots of $\theta_K^*$ 
versus missing mass and $\theta_K$ versus $Q^2$ (Figs.~\ref{fig-elastics}a 
and b, respectively).  A cut on the elastic band in the $\theta_K$ (lab
angle) versus $Q^2$ plot removes them without a significant loss of 
hyperon yield.  The $\pi^+ n$ events are removed with a simple missing-mass 
cut in which the detected hadron is assumed to be a pion.  The resulting 
hyperon missing-mass distribution over the entire kinematic range is shown 
in Fig.~\ref{mmplot}.  Both the $\Lambda(1116)$ and $\Sigma^0(1193)$ 
hyperons are apparent, along with several higher mass hyperons.
 
\subsection{Background Corrections}

\begin{figure}[tbp]
\vspace{6.2cm}
\includegraphics{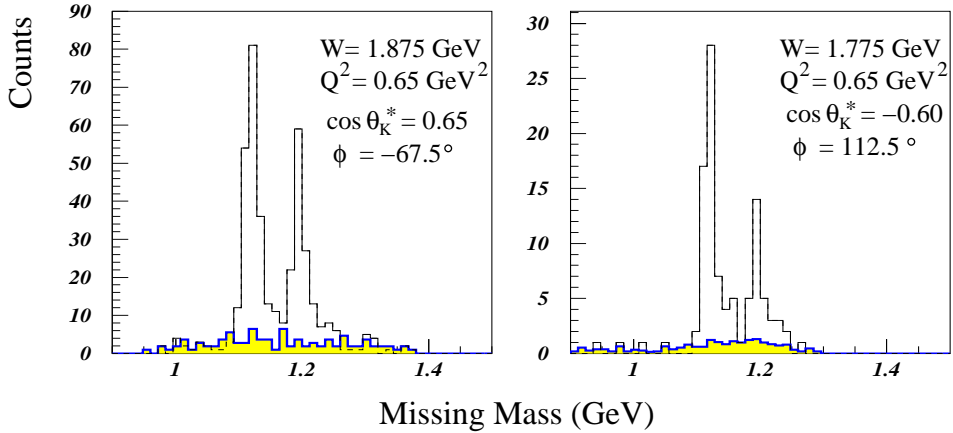}
\caption{(Color Online) Examples of the background fit results for two 
typical kinematic bins with $h=+1$.  The plots show the raw missing mass 
plots with the fitted background (solid histogram) overlaid.}
\label{fig:mmfit}
\end{figure}

To remove the multi-particle final-state background channels such as 
$e'p\pi\pi$, the phase space background was modeled by selecting the 
tails of the pion and proton mass distributions.  To do this, hadrons 
in the mass region from 0.275 to 0.725~GeV but outside of the 
momentum-dependent kaon mass cuts were selected.  Background missing 
mass distributions were calculated for these particles assigning them 
the kaon mass.  These background distributions were fit to the $p(e,e'K^+)$ 
missing mass distribution using a maximum log likelihood method appropriate 
for low statistics.  The fraction of each background distribution present 
in the data was thus estimated, and the normalized background contributions 
were subtracted from the data.  Fig.~\ref{fig:mmfit} shows the missing
mass distributions for two representative bins with the fitted background 
distributions overlaid.  The $\Lambda$ hyperon yields are the number of 
events in the background-subtracted missing mass spectra in the mass range 
from 1.095 to 1.165~GeV.

\subsection{Detector Efficiency and Acceptance}

Geometric fiducial cuts were used in order to ensure that all final
state charged particles were detected within the volume of CLAS where 
the detection efficiency is relatively large and uniform.  These cuts 
remove the edges of the CLAS detectors and depend upon the momentum of 
the particles, as well as the torus magnetic field setting.  The response 
of the CLAS detector was simulated using GSIM, a GEANT-based~\cite{geant} 
simulation package for CLAS, that combines the geometrical configuration 
with the inefficiencies of the various parts of the detector.  Monte Carlo 
techniques were used to generate $p(\vec{e},e'K^+)\Lambda$ events for each 
helicity state of the incident electrons by including the helicity-dependent
fifth structure function in the Mart and Bennhold model~\cite{mart,lee}.
   
Acceptance correction factors were obtained for each kinematic bin of
$Q^2$, $W$, $\cos\theta_K^*$, and $\phi$, and the two torus field settings,
and the effect of the acceptance corrections on the helicity-dependent
asymmetries was examined.  In the limit of large statistics in the
Monte Carlo simulation, the corrected asymmetries are indistinguishable
from the uncorrected asymmetries.  One should not expect any helicity
dependence to the CLAS acceptance outside of negligible bin-migration
effects.  Thus, the acceptance correction was observed to cancel out
(within the statistical uncertainties of both the data and Monte Carlo) 
in the asymmetry, and no acceptance corrections were applied to the 
asymmetry measurements.  However, a systematic uncertainty associated with 
not including acceptance corrections has been estimated (see
Section~\ref{sec:syserr}).

\subsection{Radiative Corrections}

Radiative corrections were performed on the extracted reaction yields 
using the exact calculation for the exclusive approach by 
Afanasev {\it et al.}~\cite{andrei}.  This approach is based on 
the covariant procedure of infrared-divergence cancellation by Bardin 
and Shumeiko~\cite{Bardin}.  The exclusive approach is used to correct
the cross section, not only in terms of the leptonic variables, but 
also the hadronic variables in exclusive electroproduction.  These 
calculations were adapted for kaon electroproduction in this work 
with a cross section model that included a contribution from 
$\sigma_{LT'}$.  The radiative correction factors were the ratio of 
the Born and the radiative cross sections, described in terms of four
kinematic variables, $Q^2$, $W$, $\cos\theta_K^*$, and $\phi$.  The 
radiative corrections are up to 30\% for a given helicity state but 
essentially cancel out in the asymmetries.

\boldmath
\subsection{Extraction of $\sigma_{LT'}$}
\unboldmath

The extraction of $\sigma_{LT'}$ requires knowledge of both the asymmetry 
$A_{LT'}$ and the unpolarized cross section $\sigma_0$, which can be seen 
by rearranging Eq.(\ref{eq:sigltp}) as:
\begin{equation}
 \label{eq:sigltp_fit}
 \frac{A_{LT'} \sigma_0}{\sqrt{\epsilon(1-\epsilon)}} = \sigma_{LT'} \sin \phi.
\end{equation}
$A_{LT'}$ is determined by forming the asymmetry of the $K^+\Lambda$ yields 
for the positive and negative beam helicity states ($h=\pm$1) as:
\begin{equation}
A_{LT'} = \frac{\frac{d\sigma}{d\Omega_K^*}^+ - \frac{d\sigma}{d\Omega_K^*}^-}
  {\frac{d\sigma}{d\Omega_K^*}^+ + \frac{d\sigma}{d\Omega_K^*}^-}  = 
  \frac{1}{P_b} \left( \frac{N^+ - N^-} {N^+ + N^-} \right),
 \label{eq:asym1}
\end{equation}
where $\frac{d\sigma}{d\Omega_K^*}^\pm$ is the cross section given by 
Eq.(\ref{eq:full_csec}) and $N^\pm$ correspond to the corrected yields 
for the positive and negative helicity states.  The electron beam is 
partially polarized, therefore, the measured asymmetries are also scaled 
by the measured beam polarization, $P_b$.  

A correction for the beam charge asymmetry (differences in the integrated 
beam charge for the different helicity states of the beam) is also included. 
This is an extremely small correction and was measured to be 
$2.99\times10^{-3}$.  It was determined by measuring the helicity-dependent 
yield ratio for $ep$ elastic scattering, which, outside of  
parity-violating effects, is exactly one.

The unpolarized differential cross sections, ${\sigma}_0$, for the
$p(e,e'K^+)\Lambda$ reaction that are used in this work, are the published 
CLAS results from the same data set~\cite{5st}.  The data from 
Ref.~\cite{5st} were bin centered in $Q^2$, $W$, and $\cos \theta_K^*$.  In 
that analysis, $\sigma_0$ was measured with the same binning in the 
variables $Q^2$, $W$, and $\cos\theta_K^*$, and the $\phi$-dependent cross 
sections were then used to extract the structure functions, 
$\sigma_U =\sigma_T + \epsilon \sigma_L$, $\sigma_{TT}$, and $\sigma_{LT}$. 

In order to smooth out the statistical fluctuations of the unpolarized
cross section data, a two-dimensional simultaneous fit in $\phi$ and
$\cos\theta_K^*$ of the data has been done.  The resulting fitted $\phi$- 
and $\cos\theta_K^*$-dependent cross sections have been used in the 
extraction of $\sigma_{LT'}$.  The measured $\phi$-dependent cross section
in a given bin $\phi^i$ is the cross section $\bar{\sigma}^i_0$ averaged
over the span of the $\phi$ bin from $\phi^i_l$ to $\phi^i_u$ (upper and
lower limits of the bin), and is given by:
\begin{eqnarray}
 \bar{\sigma}^i_0&=&\frac{1}{\Delta\phi^i} \int_{\phi^i_l}^{\phi^i_u}
 \left(\sigma_U+c_+\sigma_{LT} \cos\phi+
  \epsilon\sigma_{TT}\cos 2\phi\right)d\phi \\ \nonumber
  &=&\frac{1}{\Delta\phi^i} \left(\sigma_U\Delta\phi^i+
  c_+\sigma_{LT}\left(\sin\phi^i_u-\sin\phi^i_l\right)+
 \frac{\epsilon}{2} \sigma_{TT}\left(\sin 2\phi^i_u-\sin 2\phi^i_l\right)\right),
 \label{eq-csec2}
\end{eqnarray}
where $\Delta\phi^i=\phi^i_u-\phi^i_l$ and $c_+=\sqrt{\epsilon(1+\epsilon)}$.

In addition to the trivial $\phi$ dependence, the unpolarized cross section
has some unknown $\cos\theta_K^*$ dependence.  It has been assumed that each 
of the separated structure functions can be described by a third-order 
polynomial in $x=\cos\theta_K^*$ as:
\begin{eqnarray}
          \sigma_U &=&  U_0 + U_1x + U_2x^2 + U_3x^3, \\
    c_+\sigma_{LT} &=& LT_0 + LT_1x + LT_2x^2 + LT_3x^3,\\
\epsilon\sigma_{TT}&=& TT_0 + TT_1x + TT_2x^2 + TT_3x^3.
\label{eqn-cth1}
\end{eqnarray}

\begin{figure}[t] 
\begin{center}
\includegraphics[scale=0.45]{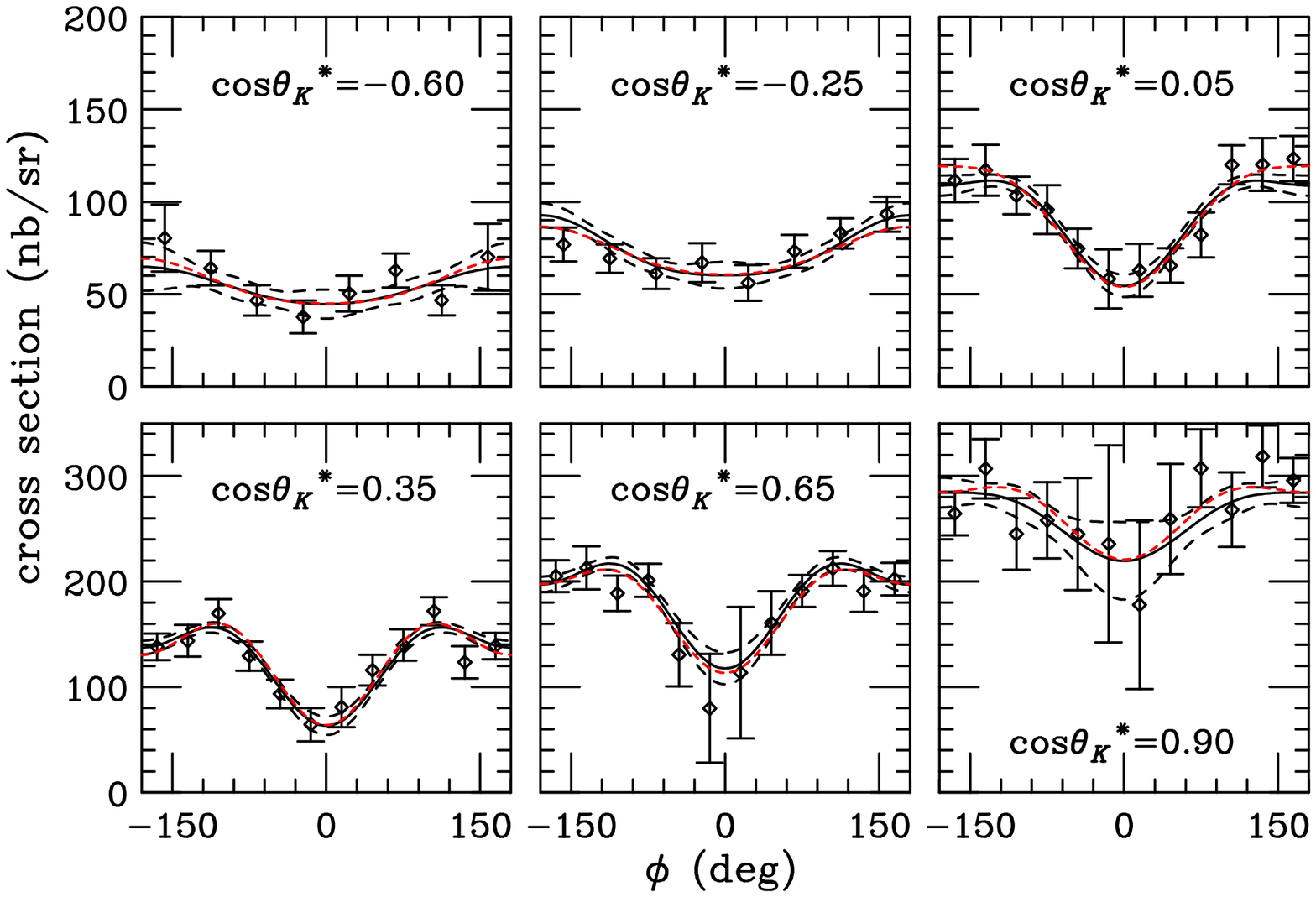}
\includegraphics[scale=0.45]{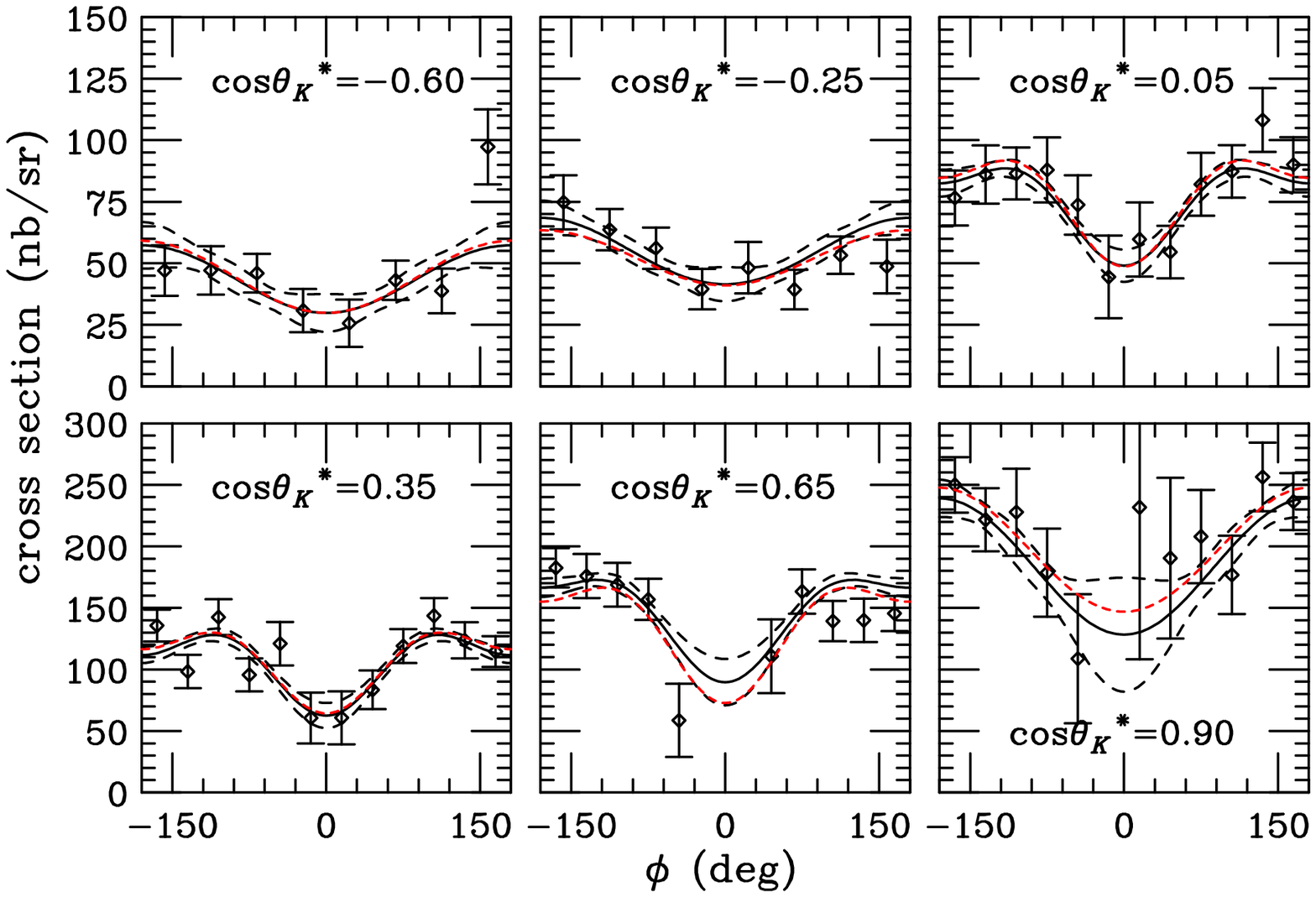}
\caption{(Color online) Fits to the unpolarized cross section $\sigma_0$ 
vs. $\phi$ for our six $\cos \theta_K^*$ points for $W$=1.875~GeV and 
$Q^2$=0.65 and 1.00~GeV$^2$ (top and bottom panels, respectively).  The 
data and red/light short-dashed curves are from Ref.~\cite{5st}.  The 
solid black curve in each plot is from the fit described in the text and 
the dashed curves represent a $\pm 1\sigma$ error band around the fit.}
\label{fig:csecs}
\end{center}
\end{figure}

Samples of the resulting fits are shown in Fig.~\ref{fig:csecs}.  In each
plot, the black solid line is the best fit and the dashed lines represent a
$\pm 1\sigma$ error band extracted from the error matrix of the fit.  As
expected, the error band is smaller than the uncertainty of the nearby data
points.  This leads to a smaller contribution to the uncertainty of
$\sigma_{LT'}$ than if the $\sigma_0$ data were used directly. The red/light
dashed lines in the figures are from using the one-dimensional $\phi$ fits
used in the structure function separation of Ref.~\cite{5st}. The
one-dimensional $\phi$ fits are very similar to the simultaneous
$\phi$/$\cos\theta_K^*$ fits and usually fall within the error band, while
the unpolarized structure functions also agree well with those extracted in
Ref.~\cite{5st}.  This parameterization of the cross section is then used
to determine the $\phi$-dependent cross section averaged over the same bin
size as each corresponding asymmetry point.

As with the cross sections, the measured asymmetries are the average values
over the span of the given $\phi$ bins.  Integrating Eq.(\ref{eq:sigltp})
over the size of the $\phi$ bin results in: 
\begin{equation}
\label{eq-AphiBC}
 A_{LT'}=A^{meas}_{LT'}\frac{\sin\phi\ \Delta\phi}{\cos\phi_l-\cos\phi_u}.
\end{equation}
The asymmetry $A_{LT'}$ has not been corrected for the finite bin size in 
the variables $Q^2$, $W$, and $\cos\theta^*_K$.  As will be discussed in
Section~\ref{sec:syserr}, such corrections are very small compared to the
uncertainties, and are very sensitive to the model choice.  Therefore, a 
systematic uncertainty associated with not making this correction has been 
estimated. 

\begin{figure}[t]
\begin{center}
\includegraphics[scale=0.45]{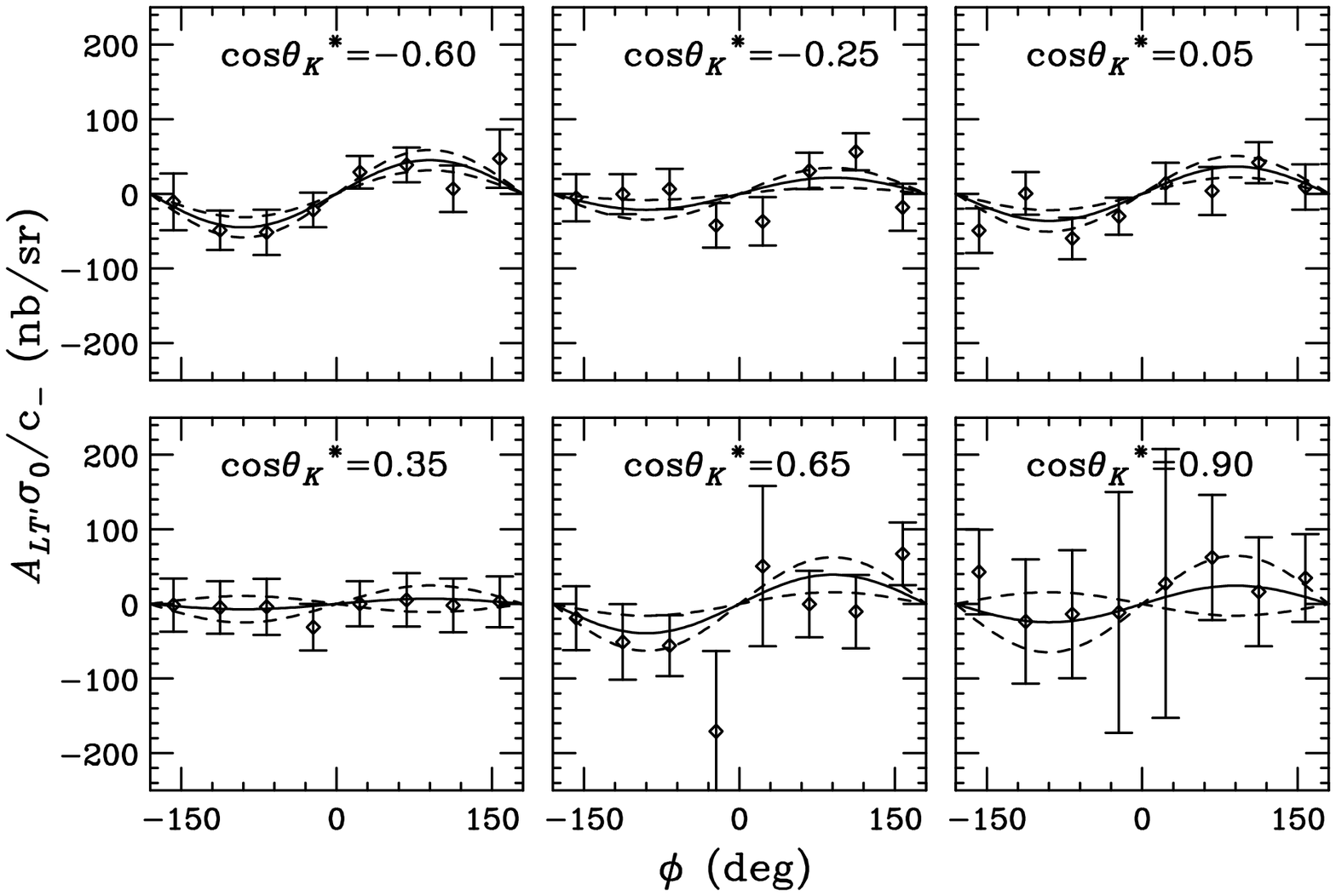}
\includegraphics[scale=0.45]{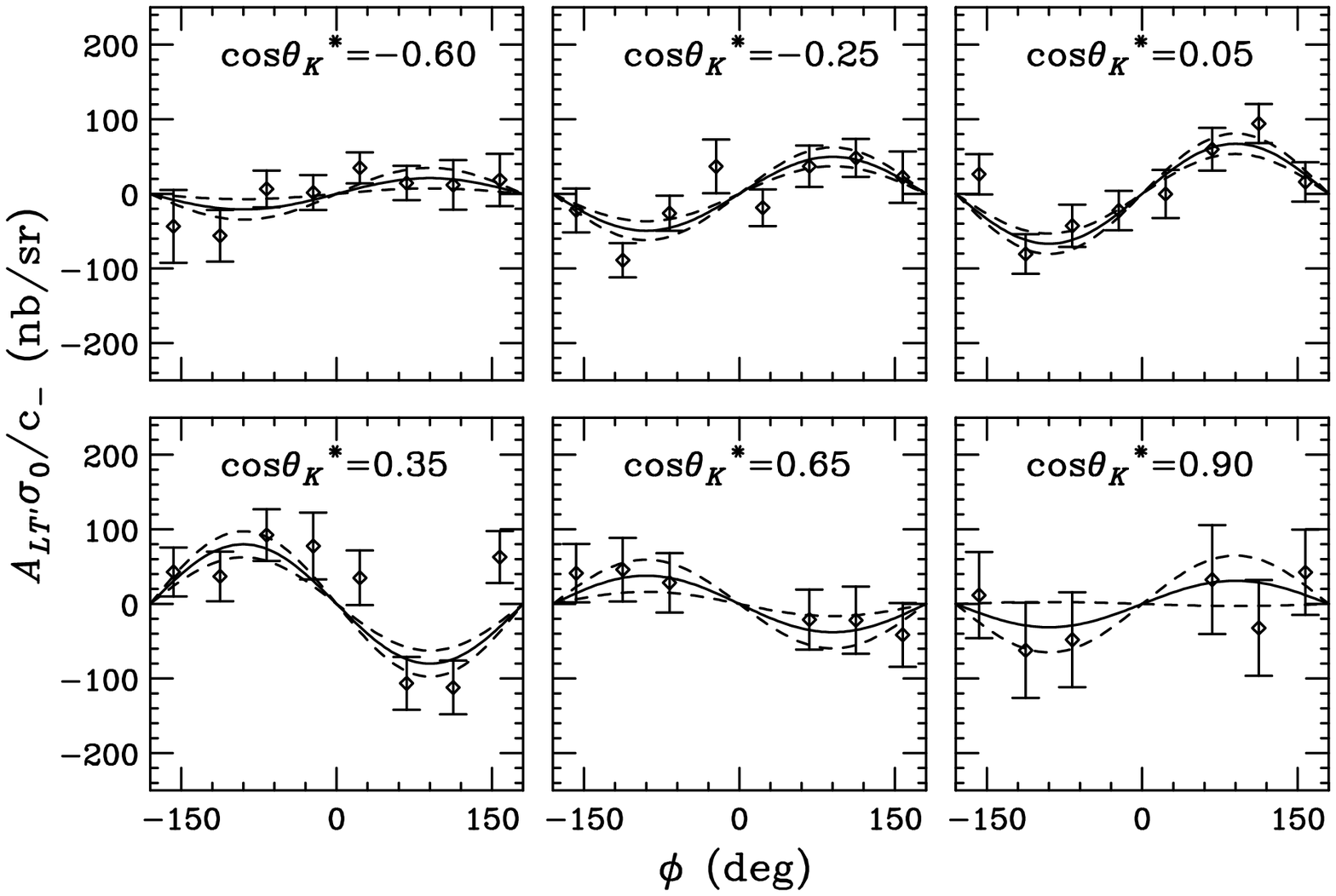}
\end{center}
\caption{Measured asymmetries multiplied by the unpolarized cross section 
and divided by the kinematic factor $c_-=\sqrt{\epsilon(1-\epsilon)}$ vs. 
$\phi$ for our six $\cos \theta_K^*$ points for a typical kinematic bin of 
$W$=1.875~GeV and $Q^2$=0.65 and 1.00~GeV$^2$ (top and bottom panels, 
respectively).  The solid curves show the results of the $\sin \phi$ fits 
and the dashed lines show the $\pm$1$\sigma$ error band from the fits.}
\label{fig:Afit}
\end{figure}

To extract $\sigma_{LT'}$, a simple sine fit was performed according to
Eq.(\ref{eq:sigltp_fit}), where the kinematic factor
$\sqrt{\epsilon(1-\epsilon)}$ has been calculated at the bin-centered
value of $Q^2$ and $W$ for each bin (see Table~\ref{tab:bins}).  Samples 
of the data and the resulting fits are shown in Fig.~\ref{fig:Afit}.  The 
solid curves are the fit result and the dashed curves indicate the 
$\pm 1 \sigma$ error band from the fit.

The error bars on the data points are a combination of the contributions
from both $A_{LT'}$ and $\sigma_0$, and are given by:
\begin{equation}
\label{eq-sgltp-fit}
\delta(A_{LT'}\sigma_0) = \sqrt{ {(A_{LT'}\delta\sigma_0)}^2 +
  {(\sigma_0 \delta A_{LT'})}^2}.
\end{equation}
The uncertainty $\delta A_{LT'}$ is the quadrature sum of the statistical 
and $\phi$-dependent systematic uncertainties (see Section~\ref{sec:syserr} 
for details), while $\delta\sigma_0$ comes from the fit of the cross 
sections described above.

\boldmath
\subsection{Systematic Uncertainties}
\label{sec:syserr}
\unboldmath

Various sources of systematic uncertainty that affect the measured
asymmetries $A_{LT'}$ and the extracted structure functions $\sigma_{LT'}$ 
are considered in this analysis.  The sources of systematic uncertainty 
that affect the measured asymmetries include uncertainties due to yield 
extraction, fiducial cuts, acceptance corrections, radiative corrections, 
and the beam charge asymmetry.  These are uncorrelated point-to-point 
uncertainties.  Scale-type uncertainties affect $\sigma_{LT'}$ only and 
include the bin centering and beam polarization uncertainties, as well as 
the systematic uncertainties in the measurement of the unpolarized cross 
section, $\sigma_0$.  Table~\ref{tab:sys} summarizes the various systematic
uncertainties that affect $A_{LT'}$ and $\sigma_{LT'}$.

In the case of the $\phi$-dependent uncertainties, all but the yield 
extraction uncertainty are dominated by statistical uncertainties.
The uncertainty due to the background-subtraction (yield extraction) 
procedure was estimated to be the same as was determined in the 
cross section extraction procedure~\cite{5st}.  In that analysis, various 
changes to the procedures were studied, such as changing the histogram bin
size in the fitting procedure and using different forms for the
background shape (e.g. using both misidentified pions and protons,
only misidentified pions, and only misidentified protons), and it was
concluded that all systematic effects get larger in direct proportion to 
the size of the statistical uncertainty.  When statistics are good (roughly
100~counts/bin), the residual systematic uncertainties are very small.
It has been determined that the remaining systematic uncertainty due to 
the yield extraction is roughly equal to 25$\%$ of the size of the 
statistical uncertainty in any given bin (defined by $Q^2$, $W$, and 
$\cos \theta_K^*$).  This uncertainty was added linearly to the statistical 
uncertainty for the helicity-dependent yields (i.e. the overall statistical 
uncertainty was increased on each yield by a factor of 1.25).

In order to estimate the uncertainties due to the fiducial cuts, acceptance, 
and radiative corrections, the corrected or nominal asymmetries were 
compared to the asymmetries that resulted from using either an alternative 
correction or cut.  The RMS width of the difference between 
the nominal and alternative asymmetries, weighted by the statistical 
uncertainty of the asymmetry, was determined, and this was used as the 
estimate of the systematic uncertainty.   For the acceptance effect, the 
difference between using no acceptance correction (nominal) and applying an 
acceptance correction was studied.  This is certainly an overestimate 
in this case, however, the uncertainty is small compared to the other 
sources of systematic uncertainty and much smaller than the statistical 
uncertainty.  For the fiducial cut uncertainty, the extent of the fiducial 
cuts was varied over a large range.  The resulting asymmetries were compared
to the nominal asymmetries.  Finally, for the radiative correction 
uncertainty, two different models were used as input to the radiative 
correction code.  It has been implicitly assumed that the correction method 
is dominated by model uncertainties.

The uncertainties for the background subtraction, fiducial cuts, acceptance, 
radiative corrections, and beam charge asymmetry, are absolute uncertainties 
and were added in quadrature to the statistical uncertainty of each 
asymmetry data point before the extraction of $\sigma_{LT'}$.  
 
\begin{table}[htbp]
\begin{center}
\begin{tabular} {|c|c|c|} \hline
Type       & Source &   Systematic Uncertainty  $A_{LT'}$\\ \hline \hline 
$A_{LT'}$ & Yield Extraction Method & 0.25$\times$stat. uncertainty of yield \\ \cline{2-3}
$\phi$ dependent & Acceptance Function (GSIM) & $0.033$	\\ \cline{2-3}
 & Fiducial Cuts & $0.027$ \\ \cline{2-3}
 & Radiative Corrections & $0.009$	\\ \cline{2-3}
 & Beam Charge Asymmetry & $1.45\times10^{-4}$\\ \hline \hline
$\sigma_{LT'}$ & Bin Centering & 5.8 nb/sr (absolute) \\ \cline{2-3}
 & Beam Polarization & $\frac{\delta P_b}{P_b}\sigma_{LT'}=0.023\sigma_{LT'}$ \\ \cline{2-3}
 & Unpolarized Cross Section & $0.124\ \sigma_{LT'}$, $0.115\ \sigma_{LT'}$\\ \hline \hline
\end{tabular}
\caption{Summary of the systematic uncertainties applied to $A_{LT'}$ and 
$\sigma_{LT'}$.  The two entries for the unpolarized cross section 
uncertainty are for the $Q^2=$0.65 and 1.00~GeV$^2$ data sets.}
\label{tab:sys}
\end{center}
\end{table}

The end result of this analysis is the extraction of the fifth
structure function, $\sigma_{LT'}$, at specific points in $Q^2$, $W$, 
and $\cos\theta_K^*$ using Eq.(\ref{eq:sigltp_fit}).  These kinematic 
points are listed in Table~\ref{tab:bins} (note that our bin ``center'' 
for the $Q^2$ bin from 0.8 to 1.3~GeV$^2$ was 1.00~GeV$^2$ as given by 
Ref.~\cite{5st}, and not the true center at $Q^2$=1.05~GeV$^2$).  The 
beam-helicity asymmetry, however, is sorted into particular {\em bins} of 
$Q^2$, $W$, and $\cos \theta_K^*$, and thus a bin-centering correction 
must be considered to extract $\sigma_{LT'}$ at specific kinematic points.  
The bin-centering correction would be applied to the binned asymmetries as:
\begin{equation}
 A_{LT'}^{BC} = A_{LT'} \left( \frac{A_{LT'}^{point}}{A_{LT'}^{avg}} 
 \right )_{model} = A_{LT'} \cdot BC,
\end{equation}
where $A_{LT'}^{BC}$ represents the bin-centered beam-helicity 
asymmetry and $A_{LT'}$ represents the bin-averaged asymmetry.  
To determine the bin-centering correction factors $BC$ for this 
analysis, a model of the CLAS acceptance in $Q^2$ vs. $W$ was developed 
to account for the partially filled bins.  The $BC$ factors necessarily 
rely on a model of $A_{LT'}$, where $A_{LT'}^{point}$ is the asymmetry 
calculated at a specific kinematic point ($Q^2$, $W$, $\cos \theta_K^*$, 
$\phi$) and $A_{LT'}^{avg}$ is the calculated bin-averaged asymmetry.  
The $BC$ factors were determined using the hadrodynamic model of Mart and 
Bennhold~\cite{mart} as a starting point.  Within the framework of this 
model, several different choices of elementary reaction models are available 
with different ingredients, such as the resonant amplitudes included, as 
well as the functional forms for the meson and baryon form factors, i.e. 
the $K^+$ form factor, the $K^+K^{*+}\gamma$ transition form factor, and 
the $\Lambda$ magnetic form factor.  The differences between the structure 
functions derived using the different models for the bin-centering 
corrections on the asymmetries are quite small and none is clearly 
preferred by the asymmetry data.  The assigned systematic uncertainty 
associated with the bin-centering corrections was chosen to be the largest 
RMS width of the $\sigma_{LT'}$ differences using the different models.

The relative systematic uncertainty due to the beam polarization
measurement for the data sets used in this analysis is estimated to be 
$\frac{\delta P}{P} = 0.023$.  The estimated uncertainty on $\sigma_{LT'}$
due to the systematic uncertainty on the beam polarization is given by:
\begin{equation}
\label{eq-atl-pbeam}
\delta \sigma_{LT'}=|A_{LT'}^{meas}| \frac{\delta P_b}{P_b^2} = 
 |\sigma_{LT'}|\frac{\delta P_b}{P_b}. 
\end{equation}

The resulting values of $\sigma_{LT'}$ also have an additional 
uncertainty associated with the systematic uncertainty in the 
measurement of the unpolarized cross section, $\sigma_0$.  The estimated 
systematic uncertainties for the cross sections are given in 
Ref.~\cite{5st}, which result in corresponding systematic uncertainties 
of 0.124$\sigma_{LT'}$ and 0.115$\sigma_{LT'}$ for $Q^2$ = 0.65 and 
1.00~GeV$^2$, respectively.  The quadrature sum of the uncertainties due 
to the bin-centering correction, beam polarization, and unpolarized cross 
section, are shown by the shaded bars on the results (see 
Section~\ref{sec:result}, Figs.~\ref{fig:sigltp_cos_lowq} to 
\ref{fig:ltp-high}).


\section{Results and Discussions}
\label{sec:result}

\begin{figure}[tbp]
\begin{center}
\includegraphics[scale=0.6]{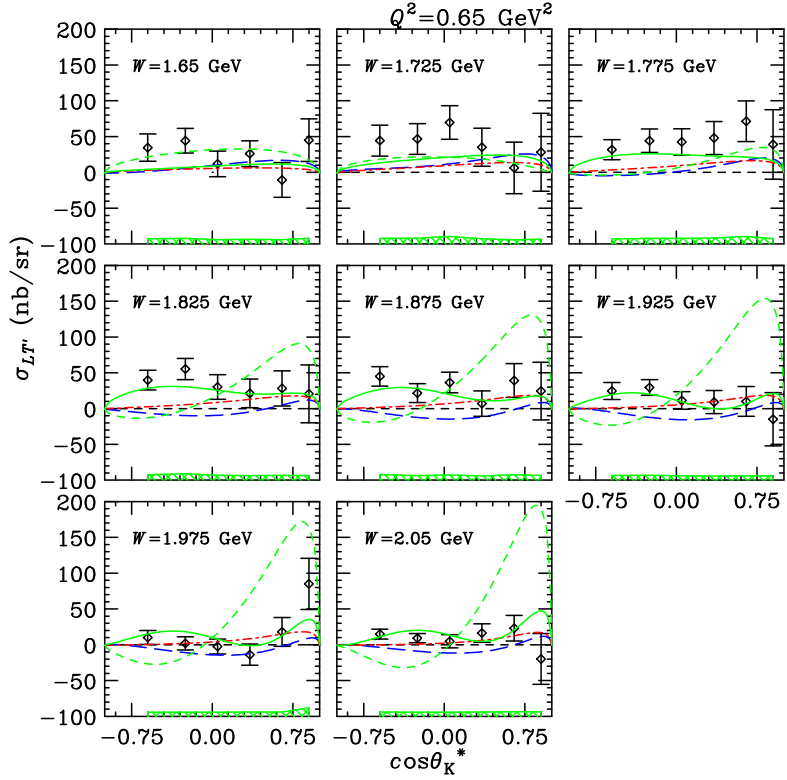}
\end{center}
\caption{(Color online) Polarized structure function $\sigma_{LT'}$ 
(nb/sr) vs. $\cos\theta_K^*$ for $Q^2$=0.65~GeV$^2$ and $W$ points as 
indicated. The curves are calculations from the MB isobar model
\cite{mart_code} (blue long dashes), the GLV Regge model~\cite{guidal_code}
(red dash-dot), the Ghent RPR model including a $D_{13}(1900)$ state
\cite{ghent_code} (green solid), and the Ghent RPR model including a 
$P_{11}(1900)$ state~\cite{ghent_code} (green short dashes).  The shaded 
bars indicate the estimated overall systematic uncertainty on the results.}
\label{fig:sigltp_cos_lowq}
\end{figure}

\begin{figure}[htbp]
\begin{center}
\includegraphics[scale=0.6]{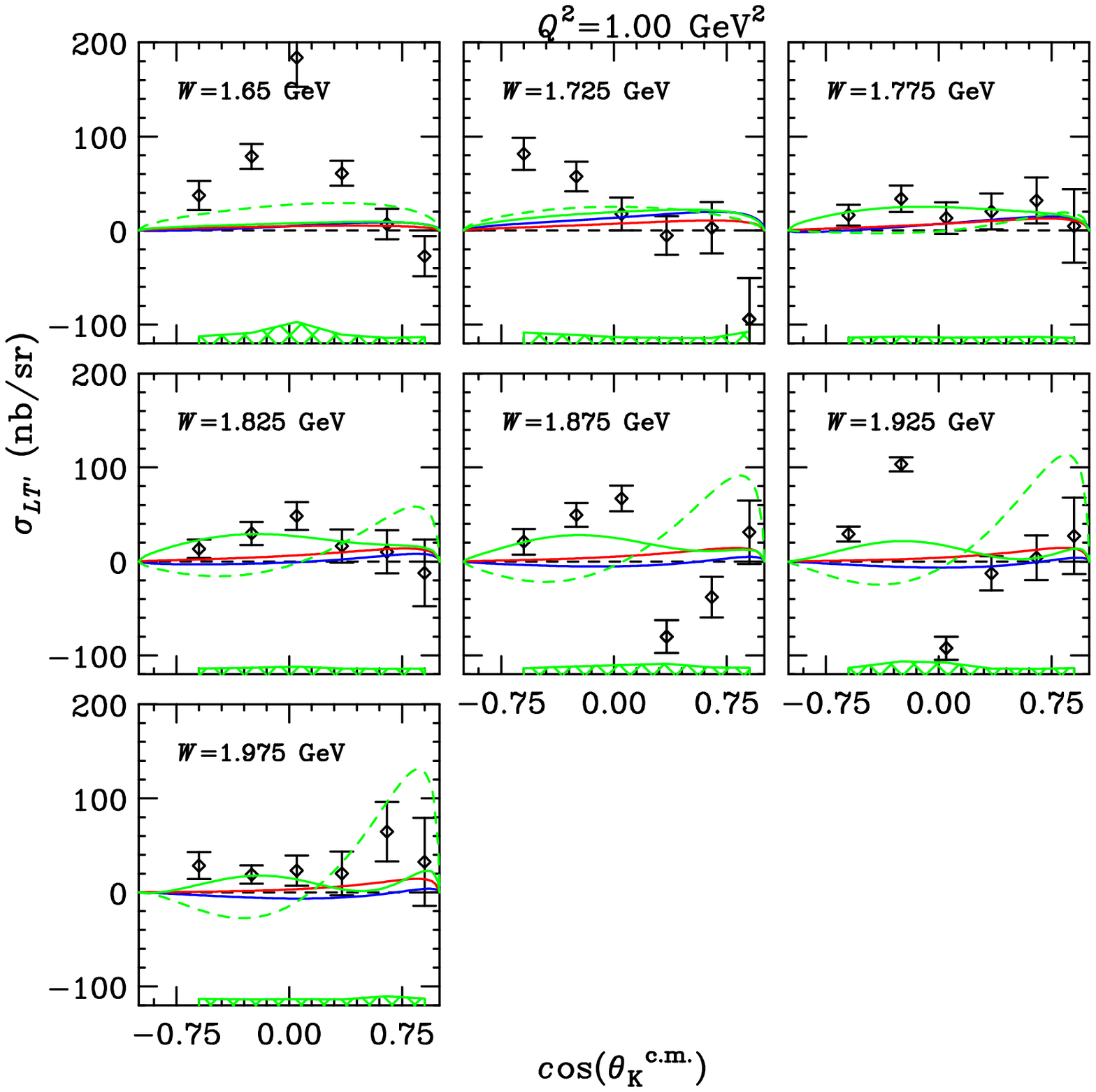}
\end{center}
\caption{(Color online) Polarized structure function $\sigma_{LT'}$ 
(nb/sr) vs. $\cos\theta_K^*$ for $Q^2$=1.00~GeV$^2$ and $W$ bins as 
indicated.  The curves are as indicated in Fig.~\ref{fig:sigltp_cos_lowq}.}
\label{fig:sigltp_cos_hiq}
\end{figure}

The angular dependence of $\sigma_{LT'}$ for various $W$ points for the
two $Q^2$ points is shown in Figs.~\ref{fig:sigltp_cos_lowq} and
\ref{fig:sigltp_cos_hiq}, along with comparisons to several model 
calculations. The lower $Q^2$ data shown in Fig.~\ref{fig:sigltp_cos_lowq} 
is rather flat over the full range of energy and angle, with no strong structures visible.  
Unlike the low $Q^2$ data, a strong $W$ and angular dependence is observed 
in the higher $Q^2$ data (Fig.~\ref{fig:sigltp_cos_hiq}).  The angular 
dependence shows an interesting peaking at middle angles for the lowest $W$ 
point ($W$=1.65~GeV), while a rapid sign change is seen at both
$W$= 1.875 and 1.925~GeV at central angles.

\begin{figure}[htbp]
\begin{center}
\includegraphics[scale=0.87]{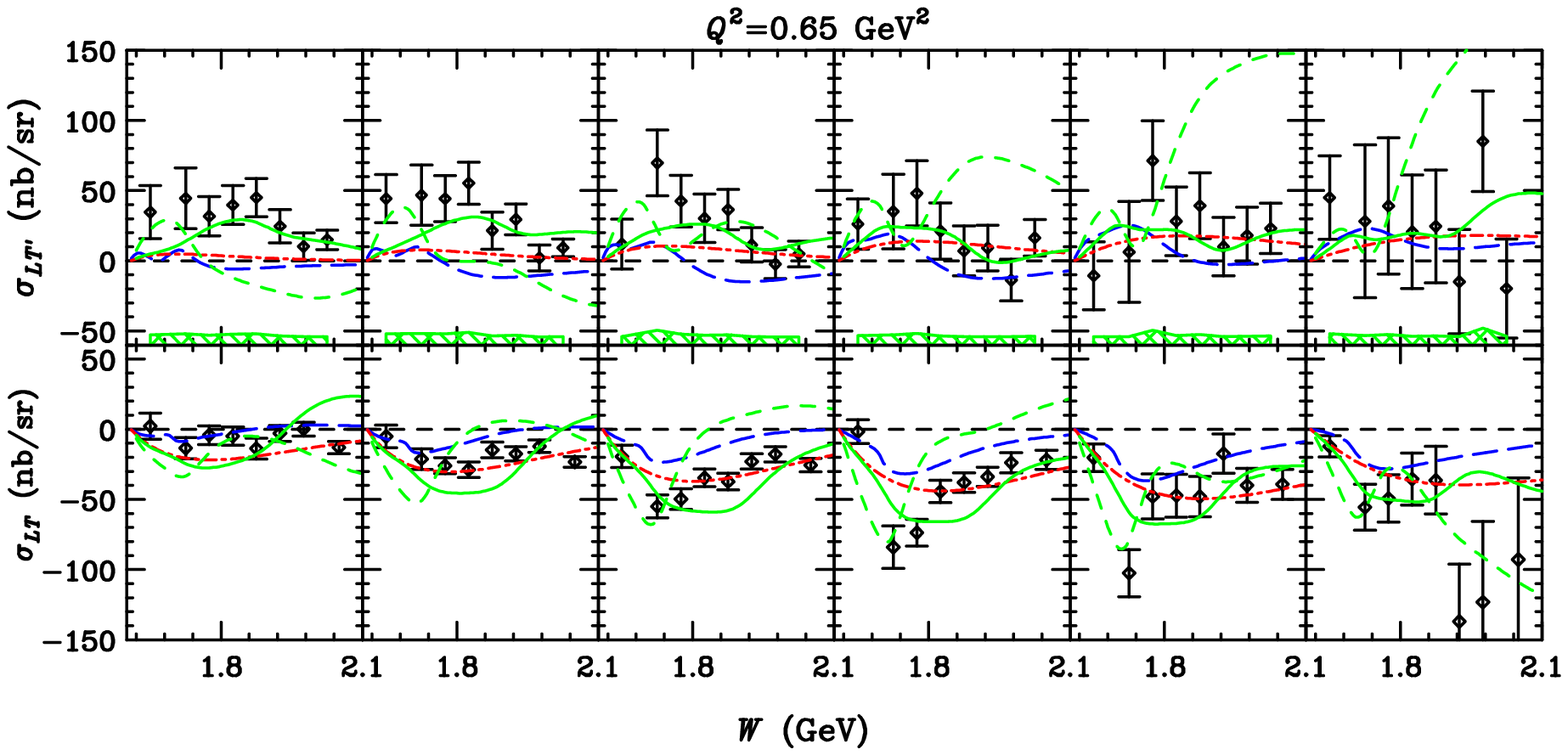}
\end{center}
\caption{(Color online) Polarized structure function $\sigma_{LT'}$ as a 
function of $W$ for various $\cos\theta_K^*$ (upper panel) compared
to the measured $\sigma_{LT}$ from Ref.~\cite{5st} (lower panel) for 
$Q^2$=0.65 GeV$^2$. Plots are shown for $\cos\theta^*_K$=-0.60, -0.25, 
0.05, 0.35, 0.65, and 0.90 (left to right).  The curves are as indicated 
in Fig.~\ref{fig:sigltp_cos_lowq}.}
\label{fig:ltp-low}
\end{figure}

\begin{figure}[htbp]
\begin{center}
\includegraphics[scale=0.87]{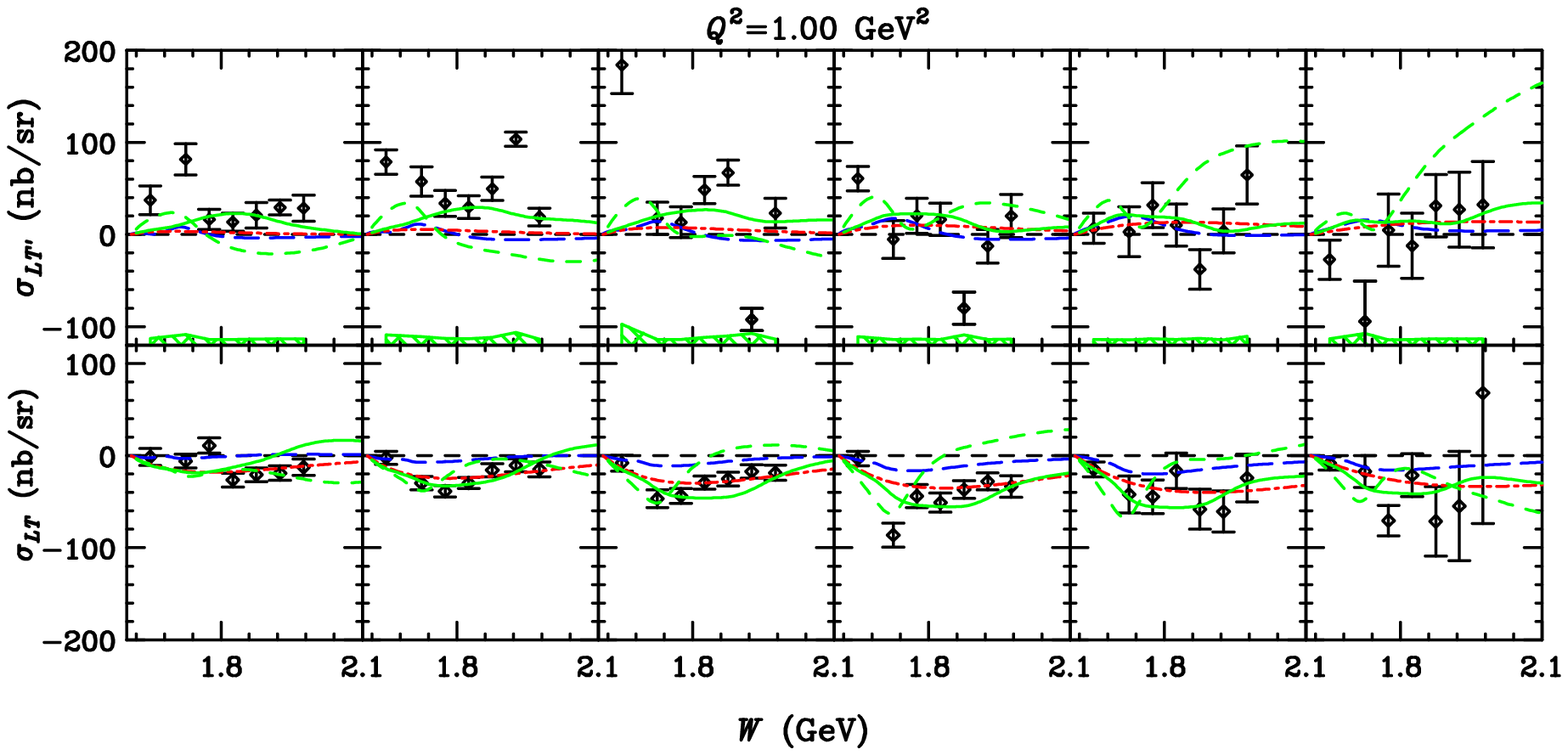}
\end{center}
\caption{(Color online) Polarized structure function $\sigma_{LT'}$ as a 
function of $W$ for various $\cos\theta_K^*$ (upper panel) compared
to the measured $\sigma_{LT}$ from Ref.~\cite{5st} (lower panel) for 
$Q^2$=1.00 GeV$^2$.  Plots are shown for $\cos\theta^*_K$=-0.60, -0.25, 
0.05, 0.35, 0.65, and 0.90 (left to right). The curves are as indicated 
in Fig.~\ref{fig:sigltp_cos_lowq}.}
\label{fig:ltp-high}
\end{figure}

The extracted $\sigma_{LT'}$ structure function results are shown as a 
function of $W$ for various $\cos\theta_K^*$ points for the $Q^2$ points 
at 0.65 and 1.00~GeV$^2$ in the top panels of Figs.~\ref{fig:ltp-low} and
\ref{fig:ltp-high}, respectively.   For the lower $Q^2$ data, 
Fig.~\ref{fig:ltp-low} shows that for the four backward-most kaon 
center-of-mass scattering angles ($\cos\theta_K^*$=-0.60, -0.25, 0.05, and
0.35), $\sigma_{LT'}$ exhibits a smooth energy dependence with a fall off
of the structure function to zero at the highest $W$ points.  In the forward 
kaon scattering angles, $\cos\theta_K^*$=0.65 and 0.90, where the reaction 
is expected to be dominated by $t$-channel exchange, $\sigma_{LT'}$ is 
consistent with zero to within the rather large error bars of the data, 
and no obvious structures are present.  This might indicate the dominance 
of a single $t$-channel exchange.

For the higher $Q^2$ data (see Fig.~\ref{fig:ltp-high}), the range of
$W$ is limited by the CLAS acceptance.  Here the $W$ dependence of 
$\sigma_{LT'}$ is similar to the lower $Q^2$ data at the more forward 
angles, $\cos\theta_K^*$=0.65 and 0.90.  However, there is a notable 
feature in the $W$ dependence in the backward and middle kaon angles. At 
$\cos\theta_K^*$=-0.25, 0.05 and 0.35, the data show an interesting 
interference feature around 1.9~GeV, with a rapid change of sign at 
$\cos\theta_K^*$=0.05 and 0.35.  While at the very backward angles, 
$\cos\theta_K^*$=-0.60, a strong enhancement is seen at about $W$=1.7~GeV
with a flat response for higher $W$.  For both of the $Q^2$ values,
$\sigma_{LT'}$ goes to zero at higher $W$.

The results are compared with calculations from the MB isobar model
\cite{mart_code} (blue long dashes), the GLV Regge model~\cite{guidal_code} 
(red dash-dot), the Ghent RPR model~\cite{ghent_code} including a 
$D_{13}(1900)$ state (green solid), and the Ghent RPR model~\cite{ghent_code}
including a $P_{11}(1900)$ state (green short dashes).  In general, none of 
the available models fully describes these data over the $Q^2$, $W$, and 
$\cos \theta_K^*$ ranges measured.  The MB and GLV models under predict the 
strength of $\sigma_{LT'}$, although they qualitatively follow the trends of 
the data.  From the comparisons, the RPR model including the $P_{11}(1900)$ 
state is clearly ruled out as already indicated in Ref.~\cite{ghent}, but 
the RPR model including the $D_{13}(1900)$ state seems to best describe the 
data qualitatively.  However, each of these models misses key features of 
the data.  The disagreements with the isobar models (MB and RPR) may not be 
too surprising as they have not been fit to these data. Therefore, the 
$\sigma_{LT'}$ structure function provides for additional new constraints 
on the model parameters.

A direct comparison of the measured polarized structure function 
$\sigma_{LT'}$ with $\sigma_{LT}$ from Ref.~\cite{5st} can reveal some 
interesting features of the data.  This is shown in Figs.~\ref{fig:ltp-low} 
and \ref{fig:ltp-high} where the polarized structure function 
$\sigma_{LT'}$ is plotted as a function of $W$ for various $\cos\theta_K^*$ 
bins and compared with $\sigma_{LT}$ at the same kinematic points.  The 
magnitudes of the two structure functions are comparable in both the lower 
and higher $Q^2$ data, although $\sigma_{LT'}$ has larger uncertainties. 
In the lower $Q^2$ data at the most backward kaon center-of-mass 
angle, $\cos\theta_K^*$=-0.60, $\sigma_{LT}$ is essentially zero, while 
$\sigma_{LT'}$ is clearly non-zero.  At $\cos\theta_K^*$=-0.25, 0.05, and 
0.35, $\sigma_{LT}$ is similar in shape and magnitude to $\sigma_{LT'}$, but 
with an opposite sign.  At the very forward kaon center-of-mass angles, 
$\cos\theta_K^*$=0.65 and 0.90, $\sigma_{LT'}$ is consistent with zero, 
while $\sigma_{LT}$ is non-zero.  In the higher $Q^2$ data, and for backward 
and middle kaon scattering angles, $\sigma_{LT'}$ has some significant 
deviations from a smooth behavior, indicating significant interferences. 
However, the shapes of $\sigma_{LT'}$ and $\sigma_{LT}$ are quite different. 

All of the data included in this work have been entered into the CLAS
physics database~\cite{database}.


\section{Conclusions}
\label{sec:conc}

The first measurements of the structure function $\sigma_{LT'}$ for 
$K^+\Lambda$ electroproduction have been reported.  The data span a range 
in $W$ from threshold to 2.05~GeV for two $Q^2$ points at 0.65 and 
1.00~GeV$^2$, and span nearly the full center-of-mass angular range of 
the final state $K^+$.  In this analysis, the energy and angular dependence 
of $\sigma_{LT'}$ have been investigated.  $\sigma_{LT'}$ is found to be
comparable in size to the unpolarized cross sections.  The structure 
function is surprisingly featureless with energy and angle for the lower 
$Q^2$ data, while the higher $Q^2$ data indicate rather strong interference 
affects near threshold and at $W$ of 1.9~GeV for central angles.  
$\sigma_{LT'}$ is consistent with zero at more forward angles and at higher 
values of $W$.

The data have been compared with several different model calculations.  The 
GLV Regge calculation generally under predicts the data.  This is perhaps not 
too surprising given that it includes no explicit $s$-channel processes, 
which are expected to show clear signatures in $\sigma_{LT'}$.  Comparisons 
to the MB isobar model and RPR hybrid isobar/Regge model (which did not
include any CLAS electroproduction data in their fits) indicate that the 
model parameters need to be tuned in order to reproduce the overall average 
strength seen in $\sigma_{LT'}$.  A bigger challenge for these models is to 
explain the strong interference signatures in the data.  Even though the
CLAS $\sigma_{LT}$ and $\sigma_{LT'}$ data have rather sizable statistical 
uncertainties, the data do have a good deal of discriminating power with 
regard to certain assumptions about which resonant states are included.

The $\sigma_{LT'}$ results were also compared with the results of the 
measurements for $\sigma_{LT}$~\cite{5st} at the same kinematic points. 
While $\sigma_{LT'}$ has larger uncertainties, the magnitudes of the 
two structure functions are comparable. In our lower $Q^2$ data, 
$\sigma_{LT'}$ is quite smooth with $W$ and $\cos \theta_K^*$.  However, at 
the high $Q^2$ value, $\sigma_{LT'}$ shows strong interference and/or FSI 
signatures at middle and backward kaon scattering angles.  Together, these
two observables provide more complete information on the amplitudes
underlying the longitudinal-transverse response for this reaction.

The question of the presence of any new resonances must wait for 
further work with the existing hadrodynamic models and partial wave 
analyses applied to the full range of the data.  Fortunately, the new 
information presented here will impose reasonable constraints on the 
amplitudes used to describe electroproduction of $K^+\Lambda$ and
$K^+\Sigma^0$ final states, making these models more reliable for future 
interpretation and prediction.

\section*{Acknowledgments}

We would like to acknowledge the outstanding efforts of the staff of 
the Accelerator and the Physics Divisions at Jefferson Lab that made 
this experiment possible.  This work was supported in part by the U.S.
Department of Energy, the National Science Foundation, the Italian 
Istituto Nazionale di Fisica Nucleare, the French Centre National de la 
Recherche Scientifique, the French Commissariat \`{a} l'Energie 
Atomique, and the Korean Science and Engineering Foundation.  The 
Southeastern Universities Research Association (SURA) operated the 
Thomas Jefferson National Accelerator Facility for the United States 
Department of Energy under contract DE-AC05-84ER40150. 


\end{document}